\documentclass[12pt, manuscript]{aastex}
\usepackage{amsmath, amsfonts, color, fullpage}
\usepackage{placeins, mathrsfs, amssymb, natbib}
\def\ltsima{$\; \buildrel < \over \sim \;$}
\def\ltsim{\lower.5ex\hbox{\ltsima}}
\def\gtsima{$\; \buildrel > \over \sim \;$}
\def\gtsim{\lower.5ex\hbox{\gtsima}}

\newcommand{\paren}[1]{\left(#1\right)}
\newcommand{\brak}[1]{\left[#1\right]}

\newcommand{\lparen}[1]{\left(#1\right.}
\newcommand{\rparen}[1]{\left.#1\right)}
\newcommand{\lbrak}[1]{\left[#1\right.}
\newcommand{\rbrak}[1]{\left.#1\right]}

\newcommand{\abs}[1]{\left|#1\right|}
\newcommand{\braket}[1]{\left<#1\right>}

\newcommand{\phivec}{\hat{\bs{\phi}}}
\newcommand{\Rvec}{\hat{\bs{R}}}
\newcommand{\zvec}{\hat{\bs{z}}}

\newcommand{\bs}[1]{\boldsymbol{#1}}

\newcommand{\parti}[2]{\frac{\partial#1}{\partial#2}}

\newcommand{\di}[2]{\frac{d#1}{d#2}}

\begin{document}

\title{The Collisionless Magnetoviscous-thermal Instability}
\author{Tanim Islam}
\affil{Lawrence Livermore National Laboratory, P.~O.~Box 808, Livermore, CA 94551-0808}
\email{islam5@llnl.gov}

\begin{abstract}
It is likely that nearly all central galactic massive and supermassive black holes are nonradiative: their accretion luminosities are orders of magnitude below what can be explained by efficient black hole accretion within their ambient environments. These objects, of which Sagittarius A* is the best-known example, are also dilute (mildly collisional to highly collisionless) and optically thin. In order for accretion to occur, magnetohydrodynamic instabilities must develop that not only transport angular momentum, but also gravitational energy generated through matter infall, outwards. A class of new magnetohydrodynamical fluid instabilities -- the magnetoviscous-thermal instability (MVTI) \citep{Islam12} -- was found to transport angular momentum and energy along magnetic field lines through large (fluid) viscosities and thermal conductivities. This paper describes the analogue to the MVTI, the collisionless magnetoviscous-thermal instability (CMVTI), that similarly transports energy and angular momentum outwards, expected to be important in describing the flow properties of hot, dilute, and radiatively inefficient accretion flows around black holes. We construct a local equilibrium for MHD stability analysis in this differentially rotating disk. We then find and characterize specific instabilities expected to be important in describing their flow properties, and show their qualitative similarities to instabilities derived using the fluid formalism. We conclude with further work needed in modeling this class of accretion flow.
\end{abstract}

\maketitle

\section{Introduction} \label{sec:intro}
Within the recent past, much progress has been made in characterizing the important dynamics of accretion flows.  The magnetorotational instability \citep{Velikhov59, Chandrasekhar60} has been applied to accretion disks \citep{Balbus91} and been shown to drive MHD fluid turbulence that can provide an outward angular momentum flux and mass accretion rate consistent with astrophysical observations, as demonstrated in a variety of numerical simulations \citep{Hawley96,  Wardle99, Sano02, DeVilliers03a, Fromang04a}. However, there exists observational evidence of hot dilute flows, in accretion about dim mass-starved supermassive black holes, for which the mean free path is of the order of the system scale or larger. Chandra X-ray observations by \cite{Baganoff03} have resolved the inner 1'' around the Sagittarius A central black hole and demonstrated that the ion mean free path at its capture radius is only a few times smaller than the system scale. The unambiguous detection of Faraday rotation in the high-
frequency radio emission about Sagittarius A \citep{Aitken00, Bower03, Marrone05} implies that the magnetic field is very easily strong enough to result in a gyrokinetic reduction in plasma dynamics. Estimates of mass accretion from the ambient conditions about this object overestimate its bolometric luminosity by approximately five orders of magnitude over radiatively efficient accretion \citep{Narayan02b}, implying that very little of the gravitational energy produced by mass accretion is radiated. However, recent nonlinear local simulations, with limits on electron pressure anisotropy due to gyrokinetic electron instabilities, show that the nonlinear development of the collisionless MRI can turbulently heat electrons sufficiently to allow these flows to become radiative \citep{Sharma07}; accretion in collisionless environments, such as those around Sag.~A*, may naturally be radiative enough that the accretion rate must remain orders of magnitude below the Bondi rate in order to explain their low luminosity. Regardless of whether this accretion is radiatively inefficient, it is very likely that in the steady state these plasmas are dilute, optically thin, and the bulk of their thermal energy lies with the protons. Furthermore, MHD plasma turbulence that transports energy generated from accretion may play an important role even in high-energy radiative, but collisionless, accretion flows.

The fact that, very plausibly, these systems may be radiatively inefficient points to the fact that these high energy, dilute plasmas are at least partially pressure supported; this is in contrast to a large class of models of radiatively efficient classical accretion disks, in which the accreting disk of matter remains geometrically thin and rotationally supported due to the efficient radiation of energy perpendicular to the disk. Numerical simulations of the MRI in a canonical black hole accretion flow \citep{DeVilliers03a, DeVilliers03b} tend to stabilize into thick disks.  The large aspect ratio of these disks invites an analysis of these disks with vertical disk structure included, or as a beginning a local analysis in which dynamically important gradients of temperature and pressure govern the nature of local instabilities.

A formulation of magnetized plasma dynamics that is especially well-suited for collisionless or mildly collisional MHD plasma equilibrium and dynamics is that of Kulsrud's drift-kinetic approximation to the Boltzmann equation \citep{Kulsrud83, Kulsrud05}. To lowest order the particle distribution function is characterized by dynamics only along magnetic field lines, MHD conditions of quasi-neutrality with ions and electrons moving together, and conservation of magnetic moment for particle distributions. Furthermore, it is expected that additional dynamics that cannot be modeled through the Kulsrud formalism, such as momentum and energy transfer processes resulting in temperature equilibration or electric resistivity, may not be dynamically important to a first approximation.

For this problem we consider the following hierarchy of scales appropriate to lowest-order gyro kinetic expansion: $1/T < \omega_{pi} \ll \Omega_{ci}$, $1/L < \omega_{pi}/c \ll \rho_i$, where $\omega_{pi}$ is the ion plasma frequency, $\Omega_{ci}$ is the ion gyrofrequency, $\rho_i$ is the ion gyroperiod, $\omega_{pi}/c$ is the inverse ion inertial depth, and $L$ and $T$ are the shortest length and fastest time scales associated with this system. Densities are large enough that Alfv\'{e}n velocities are smaller than the speed of light, therefore relativistic MHD effects may be ignored. The gravitational acceleration is purely due to that of the central object. We consider a plasma equilibrium where pressures parallel and perpendicular to the magnetic field are equal, hence the equilibrium particle distribution for electrons and ions has one temperature. We formulate the problem in a cylindrical geometry, where the axis of rotational lies along the vertical axis. $\Rvec$, $\phivec$, and $\zvec$ are unit vectors in the radial, azimuthal, and vertical directions, respectively.

The organization of this paper is as follows: in \S\ref{sec:nomenclature} we discuss the variables and nomenclature used in this paper; in \S\ref{sec:drift-kinetic}, we use a form of the drift kinetic equation that represents particle dynamics in a co-rotating frame, explicitly state the equilibrium we choose in our local analysis, and include total MHD force balance and MHD induction equations in a co-rotating frame. In \S\ref{sec:dilute_fluxes} we justify and modify turbulent and average wave quantities appropriate to characterize accretion (see, e.g., \cite{Balbus98, Balbus04a}) in dilute and radiatively inefficient magnetized flows. In \S\ref{sec:stability-analysis} we consider the stability of hot dilute rotating plasmas to a new instability, the collisionless analogue to the magnetoviscous-thermal instability (MVTI) \citep{Islam12}, the collisionless MVTI or CMVTI. We also demonstrate that quadratic estimates of heat fluxes and Reynolds stress are of the right form to drive accretion in this dilute thick flow. In \S\ref{sec:summary_and_future_work} we summarize our main results as well as describe directions for further research.

\section{Variables and Nomenclature} \label{sec:nomenclature}
Our coordinate system for the rotating disk is a cylindrical system located about the central mass. $R$ is the radial coordinate, $\phi$ is the azimuthal angle, and $z$ the vertical coordinate aligned along the axis of rotation. $\paren{\Rvec, \phivec, \zvec}$ refer to unit vectors in the radial, azimuthal, and vertical directions, respectively. For field variables of temperature $T$, pressure $p$, density $\rho$, electric and magnetic fields ${\bf E}$ and ${\bf B}$, and pressure $p$, we use the following notation:
\begin{itemize}
	\item Equilibrium value of, say density: $\rho_0$,
	\item Perturbed density: $\delta\rho$,
	\item Total density (equilibrium + perturbed): $\rho = \rho_0 + \delta\rho$.
\end{itemize}
For velocity, we use the following notation:
\begin{itemize}
	\item Primary equilibrium flow velocity, which is azimuthal: ${\bf V}_0 = R\Omega(R)\phivec$, where $\Omega(R)$ is the orbital angular velocity,
	\item Perturbed flow velocity: $\delta{\bf u}$,
	\item Total flow velocity: ${\bf V} = R\Omega(R)\phivec + {\bf u}$.
\end{itemize}
Components of an equilibrium vector quantity, such as the radial component of the equilibrium magnetic field, are written as $B_{R0}$. The radial component of, for instance, the perturbed magnetic field is denoted as $\delta B_R$.

In this paper, we consider an electron-ion plasma. In these systems, the ions and electrons are only very weakly coupled collisionally, so the pressure of both species may differ significantly. In an electron-ion plasma, $p_i$ refers to ion pressure, $p_e$ to electron pressure. Quantities such as equilibrium ion pressure will be denoted as $p_0$, while perturbed species variables such as perturbed ion pressure are denoted as $\delta p_i$. Finally, in a collisionless MHD plasma we find it convenient to consider fluid properties, such as parallel $p_{\|}$ and perpendicular $p_{\perp}$ pressure, that are velocity moments of the particle distribution function. For example, the equilibrium ion parallel pressure is defined as $p_{i\| 0}$ and the perturbed ion parallel pressure is $\delta p_{i\|}$.

\section{The Drift Kinetic And Constituent Equations in Rotating Frame}
\label{sec:drift-kinetic}
In this section we state the equations and disk equilibrium used in the eigenmodal analysis and the demonstration of quadratic heat flux of the CMVTI. Without derivation (see, e.g., \cite{Hinton76, Sharma03, Sharma06}), the collisionless drift kinetic equation in a rotating frame can be shown to be of the following form,
\begin{eqnarray}
  &&\begin{aligned}{}
    &\paren{\parti{}{t} + \Omega\parti{}{\phi}}\paren{f_s B} + \nabla\cdot\paren{\brak{v_{\|}{\bf b} + {\bf u}_{\perp}}f_s B} + \parti{}{v_{\|}}\paren{f_s B\brak{\frac{Z_s e}{m_s}E_{\|} + \frac{1}{m_s n_0}{\bf b}\cdot\nabla p_{s0}}} + \\
    &\parti{}{v_{\|}}\lparen{f_s B\lbrak{-{\bf b}\cdot\paren{\brak{\parti{}{t} + \Omega\parti{}{\phi}}{\bf u}_{\perp} + \brak{v_{\|}{\bf b} + {\bf u}_{\perp}}\cdot\nabla{\bf u}_{\perp}} + \mu B\nabla\cdot{\bf b} +}} \\
    &\rparen{\rbrak{2\Omega\zvec\cdot\paren{{\bf b}\times{\bf u}} - b_{\phi} R\paren{{\bf u}_{\perp} + v_{\|}{\bf b}}\cdot\nabla\Omega}} = 0,
  \end{aligned} \label{eq:drift-kinetic}
\end{eqnarray}
$f_s$ is the species particle distribution function, and $m_s$ and $Z_s$ is the mass and charge of a particle of species $s$. $v_{\|}$ is the component of corotating velocity along the magnetic field, and $\mu$ is the magnetic moment ($m v_{\perp}^2 / \paren{2B}$). Additional terms appear in the formulation of Eq.~(\ref{eq:drift-kinetic}) that do not appear explicitly in the normal drift-kinetic equation of \cite{Kulsrud83}: non-inertial rotational accelerations along the magnetic field, $2\Omega\zvec\cdot\paren{{\bf b}\times{\bf u}} - b_{\phi}R\paren{{\bf u}_{\perp} +  v_{\|}{\bf b}}\cdot\nabla\Omega$, and accelerations along the magnetic field associated with large thermal energies, $1/\paren{m_s n_0}{\bf b}\cdot\nabla p_{s0}$. 

Next, the form of the full MHD force balance and induction equations in a co-rotating frame are given by,
\begin{eqnarray}
	&&\begin{aligned}
      &\rho\paren{\brak{\parti{}{t} + \Omega\parti{}{\phi}}{\bf u} +
        {\bf u}\cdot\nabla{\bf u} - 2\Omega{\bf u}\times\zvec + R {\bf u}\cdot\nabla\Omega\phivec} =
      \frac{1}{c}{\bf J}\times{\bf B} +
      \frac{n}{n_0}\nabla p - \nabla\cdot\mathbb{P},
    \end{aligned} \label{eq:MHD_force_balance} \\
	&&\begin{aligned}{}
		&\parti{\bf B}{t} = -{\bf u}\cdot\nabla{\bf B} - {\bf B}\paren{\nabla\cdot{\bf u}} + {\bf B}\cdot\nabla{\bf u} + R\phivec{\bf B}\cdot\nabla\Omega - \Omega\parti{\bf B}{\phi}.
    \end{aligned} \label{eq:induct_total}
\end{eqnarray}
Where $p = p_{e} + p_{i}$, $\mathbb{P} = p_{\perp}\mathbb{I} + \paren{p_{\|} - p_{\perp}}{\bf b}{\bf b}$,
$p_{\|} = p_{i\|} + p_{e\|}$, and $p_{\perp} = p_{i\perp} + p_{e\perp}$. Parallel, perpendicular, and total pressures are given by their standard forms,
\begin{eqnarray}
	&&p_{s\|} = 2\pi\int m_s \paren{v_{\|} - u_{\|}}^2 f_s \paren{B\,d\mu\,dv_{\|}}, \\
	&&p_{s\perp} = 2\pi\int m_s \mu B f_s\paren{B\,d\mu\,dv_{\|}}, \\
	&&p_s =\frac{2}{3} p_{s\perp} + \frac{1}{3} p_{s\|}. \label{eq:total_pressure}
\end{eqnarray}
In this paper we analyze the stability and quadratic transport of an equilibrium geometrically thin nonradiative, collisionless disk at its midplane. For simplicity, temperature is independent of height above the disk. To lowest order there is no net current, the plasma velocity is purely azimuthal, and the equilibrium magnetic field is nonradial and axisymmetric. Therefore, the electron and ion pressure and density as a function of vertical coordinate $z$ goes as,
\begin{eqnarray}
	&&n_{i0}, n_{e0}, p_{i0}, p_{e0} \sim \exp\paren{-\frac{z^2}{2H^2}}. \label{eq:disk_density_pressure_vertical_profile}
\end{eqnarray}
The disk scale height $H$ is given by,
\begin{eqnarray}
	&&H^2 = \frac{k_B \paren{T_{i0} + T_{e0}}}{\paren{m_i + m_e} \Omega^2}. \label{eq:disk_scale_height}
\end{eqnarray}
The equilibrium solution to Eqs~(\ref{eq:drift-kinetic}) is,
\begin{eqnarray}
  &&f_{s0} = \frac{n_0(z = 0)}{\paren{2\pi k_B  T_{s0}/m_s}^{3/2}} \exp\paren{-\frac{z^2}{2H^2} - \frac{m_s v_{\|}^2}{2k_B T_{s0}} - \frac{m_s\mu B}{k_B T_{s0}}}. \label{eq:equilibrium_distribution_function}
\end{eqnarray}
The equilibrium magnetic field ${\bf B}_0$ and its vector normal ${\bf b}_0$ are,
\begin{eqnarray}
	&&{\bf B}_0 = B_0 \paren{\phivec \sin\chi + \zvec \cos\chi }, \label{eq:equlibrium_magnetic_field} \\
	&&{\bf b}_0 = \phivec \sin\chi + \zvec \cos\chi. \label{eq:equilibrium_magnetic_normal}
\end{eqnarray}

Global equilibria of axisymmetric, and at least partially rotationally supported, plasmas \citep{Hinton76, Bisnovatyi-Kogan85, Ogilvie97} are characterized by a complicated global geometry due to the requirements of centrifugal force balance and equilibrium along axisymmetric magnetic surfaces. Local analysis away from the disk midplane, or global analysis of the longer wavelength CMVTI in a high-aspect ratio collisionless accretion disk, is beyond the scope of this paper.

\section{Turbulent and Wave Fluxes For Dilute Rotating Plasmas} \label{sec:dilute_fluxes}
The evolution equation for the total energy within a disk, using methods outlined in \cite{Balbus98}, is given by the following (see, e.g., \cite{Sharma06}):
\begin{eqnarray}
  &&\begin{aligned}
      &\paren{\parti{}{t} + \Omega\parti{}{\phi}}\paren{\frac{1}{2}\rho u^2 + \frac{3}{2}p + \frac{B^2}{8\pi}} +
      \nabla\cdot{\mathcal F}_E - \rho{\bf u}\cdot\frac{1}{\rho_0}\nabla p_0 = \\
      &-\parti{\Omega}{\ln R} W_{R\phi} - R\parti{\Omega}{z} W_{z\phi} - Q_-.
  \end{aligned} \label{eq:energy_balance}
\end{eqnarray}
A fuller derivation of Eq.~(\ref{eq:energy_balance}) can be found in, e.g., \cite{Islam07}.
${\mathcal F}_E$ is the heat flux arising from local fluctuations, $W_{R\phi}$ is the azimuthal stress, $W_{z\phi}$ is the vertical-azimuthal stress, $Q_-$ is a radiative loss term. One may look to \cite{Sharma06, Islam07}, for fuller derivations of the energy balance term including the pressure expression term. In the context of disk accretion theory, the above expresses the fact that energy is generated by azimuthal stresses that couple to the free energy available from radial and vertical angular velocity gradients. This energy can then be accounted for in various ways: in a classical accretion disk, the energy flux is almost wholly radiated away; in a geometrically thick accretion disk, turbulent heat fluxes are large enough to transport at least some of this viscously generated energy
\citep{Balbus98, Balbus03}. Even in collisionless accretion, turbulent energy generation may heat electrons until they become radiatively efficient at locally dissipating energy \citep{Sharma07}. However, in nonradiative flows \citep{Narayan98a}, viscously generated energy must be carried away by a turbulent heat flux \citep{Balbus04a}.

The energy flux is given by,
\begin{eqnarray}
  &&\begin{aligned}
      &{\mathcal F}_E = {\bf u}\paren{\frac{1}{2}\rho u^2 +
        \frac{5}{2}p} + \frac{1}{4\pi}{\bf
        B}\times\paren{{\bf u}\times{\bf B}} + {\bf b}q + p_v\paren{\brak{{\bf u}\cdot{\bf b}}{\bf b} - \frac{1}{3}{\bf u}}.
    \end{aligned} \label{eq:energy_flux}
\end{eqnarray}
The first term in ${\mathcal F}_E$ corresponds to flux of gas kinetic energy, the second to the enthalpy, and the third term corresponds to Poynting MHD flux. The heat flux $q = q_i + q_e$ and pressure difference $p_v = p_{vi} + p_{ve}$ are defined in, e.g., \cite{Chang92a, Chang92b} in the context of heat flux expressions to model collisionless transport due to specific instabilities into a fluid formalism, and shown here,
\begin{eqnarray}
	&&\begin{aligned}{}
	   &q_s = \frac{1}{2}q_{s\|} + q_{s\perp},\\
	   &q_{s\|} = 2\pi \int m_s\paren{v_{\|} - u_{\|}}^3 \paren{B\,d\mu\,dv_{\|}} f_s,\\
	   &q_{s\perp} = 2\pi \int m_s\paren{v_{\|} - u_{\|}}\paren{\mu B}\paren{B\,d\mu,dv_{\|}} f_s.
	\end{aligned} \label{eq:heat_flux} \\
	&&\begin{aligned}
		&p_{sv} = p_{s\|} - p_{s\perp}.
	\end{aligned} \label{eq:pressure_difference}
\end{eqnarray}
The fourth and fifth terms of Eq.(~\ref{eq:energy_flux}) correspond to contributions due to heat fluxes along the magnetic field and the viscous stress. $W_{R\phi}$ and $W_{z\phi}$ are given by,
\begin{eqnarray}
  &&W_{R\phi} = \rho u_R u_{\phi} - \frac{B_R B_{\phi}}{4\pi}
  + p_v b_R b_{\phi}, \label{eq:azimuthal_stress} \\
  &&W_{z\phi} = \rho u_z u_{\phi} - \frac{B_z B_{\phi}}{4\pi}
  + p_v b_z b_{\phi}. \label{eq:vertical-azimuthal_stress}
\end{eqnarray}
The angular momentum flux can be derived from MHD force balance and continuity, and for an accretion disk is given by \citep{Balbus98, Islam07},
\begin{eqnarray}
  &&\begin{aligned}
      &\paren{\parti{}{t} + \Omega\parti{}{\phi}}\paren{\rho R \brak{u_{\phi} + R\Omega}} + \\
      &\nabla\cdot R\paren{\rho{\bf u}\brak{u_{\phi} + R\Omega} -
        \frac{B_{\phi}{\bf B}}{4\pi} + p_v b_{\phi}{\bf b} + \brak{p_{\perp} +
          \frac{B^2}{8\pi}}\phivec} = 0. \end{aligned} \label{eq:azimuthal_evolution}
\end{eqnarray}
To understand how local fluctuations about mean quantities of the form
$A = A_0 + \delta A$, whether waves or turbulence, can tap into
sources of energy within this rotating system, it is easiest to
consider the truncated dynamics of this system by averaging vertically
and azimuthally. Define the following averaged quantity:
\begin{eqnarray}
  &&\braket{A} = \frac{1}{H}\int_0^{2\pi}\int_{z=-\infty}^{z=\infty} A\,dz\,d\phi,
\end{eqnarray}
and consider fluctuations which spatially average to zero, i.e. $\braket{\delta A} = 0$. Contributions of fluctuations appear at second order. Since in equilibrium ${\bf u}_0 = {\bf 0}$, $p_{\| 0} = p_{\perp 0} = p_0$, $q_0 = 0$, and $q_{v,0} = 0$, the energy and angular momentum equations are,
\begin{eqnarray}
  &&\parti{\braket{L}}{t} + \frac{1}{R}\parti{}{R}\paren{R^3\Omega
    \braket{\rho u_R} + R\braket{W_{R\phi}}} =
  0, \label{eq:averaged_angular_momentum_equation} \\
  &&\begin{aligned}
      &\parti{\braket{\mathcal E}}{t} +
      \frac{1}{R}\parti{}{R}R\braket{{\mathcal F}_{ER}} -
      \braket{\rho u_R}\frac{1}{\rho_0}\parti{p_0}{R} = -\parti{\Omega}{\ln
        R}\braket{W_{R\phi}} - Q_-.
    \end{aligned} \label{eq:averaged_total_energy_equation}
\end{eqnarray}
We have ignored the flux of gas kinetic energy, that appears at third order in fluctuating quantities, and the Poynting flux, which is subdominant to the other terms in the energy flux. We have taken $W_{z\phi}$ to be an even function of height. From Eq.~(\ref{eq:equilibrium_magnetic_normal}), the equilibrium azimuthal component of the magnetic normal vector is $\cos\chi$.
\begin{eqnarray}
  &&\braket{L} = \braket{\rho R\paren{u_{\phi} +
      R\Omega}}, \label{eq:angular_momentum_average} \\
  &&\braket{\mathcal E} = \braket{\frac{1}{2}\rho u^2 +
    \frac{1}{2}p_{\|} + p_{\perp} +
    \frac{B^2}{8\pi}}, \label{eq:energy_average} \\
  &&\braket{W_{R\phi}} = \braket{\rho_0\delta u_R \delta u_{\phi} -
    \frac{\delta B_R \delta B_{\phi}}{4\pi} + \delta p_v \delta b_R \sin\chi}, \label{eq:azimuthal_stress_average} \\
  &&\braket{F_{ER}} = \frac{5}{2}\rho_0\braket{\delta u_R\delta\theta}
  + \braket{\delta q \delta b_R} - \frac{1}{3}\braket{\delta p_v
    \delta u_R}. \label{eq:averaged_radial_heat_flux}
\end{eqnarray}
Note that the radial mass flux term $\braket{\rho u_R} =
\braket{\delta\rho \delta u_R} + \rho_0 u_{R2}$, where $u_{R2}$ is a
second order steady bulk radial flow of matter with magnitude of order
$\abs{\delta\rho/\rho_0}^2$; as noted by \cite{Balbus03}, in a steady-state geometrically thin disk, the net radial matter flux has a magnitude given by $\abs{\braket{\rho u_R}} \sim \rho \abs{\braket{{\bf u}}^2}/\paren{R\Omega}$.

\section{Stability Analysis And Quadratic Heat Fluxes} \label{sec:stability-analysis}
The discussion of the CMVTI is divided into the following subsections. \S\ref{subsec:perturbed_equations} and \ref{subsec:perturbed_pressure_disp_relation} describe the eigenmodal equations, and collisionless pressure expressions, used to derive the full dispersion relation for the CMVTI, which is not shown in this work. In the limit of zero equilibrium pressure and temperature gradients, the CMVTI reduces to the collisionless MRI. \S\ref{subsec:quadraticfluxes} estimates quadratic modal expressions for heat flux and Reynolds stress from the CMVTI. We find it useful to use the following variables and scalings:
\begin{eqnarray}
  &&\begin{aligned}
  	&\theta_0 = \frac{k_B T_0}{m_i} \\
  	&v_A^2 = \frac{B_0^2}{4\pi\rho_0} \\
  	&x = k_{\|} v_A/\Omega \\
  	&\hat{\bf k} = {\bf k} v_A/\Omega \\
  	&\gamma = \Gamma/\Omega \\
  	&\alpha_P = -\paren{\theta_0^{1/2}/\Omega}\parti{\ln p_0}{R} \\
  	&\alpha_T = -\paren{\theta_0^{1/2}/\Omega}\parti{\ln T_0}{R} \\
  	&\beta = \theta_0/v_A^2.
  \end{aligned} \label{eq:normalization_variables}
\end{eqnarray}
Our expression for the Alfv\'{e}n speed $v_A$ differs by a factor of $\sqrt{2}$ from the standard definition. We also explore the stability of stratified media that are convectively stable, hence one in which $\alpha_S < 0$ or equivalently $\alpha_T < \frac{2}{5}\alpha_P$. All plots of dispersion relations, heat fluxes, and Reynolds stresses use a plasma equilibrium with $\chi = \pi/4$ (equal equilibrium toroidal and vertical magnetic field components), plasma $\beta = 10^2$, Keplerian rotation profile $\Omega \propto R^{-3/2}$, and purely vertical wavenumbers. 

\subsection{Perturbed Axisymmetric Distribution Function at the Mid-plane} \label{subsec:perturbed_equations}
\begin{figure}
	\includegraphics[width=\linewidth]{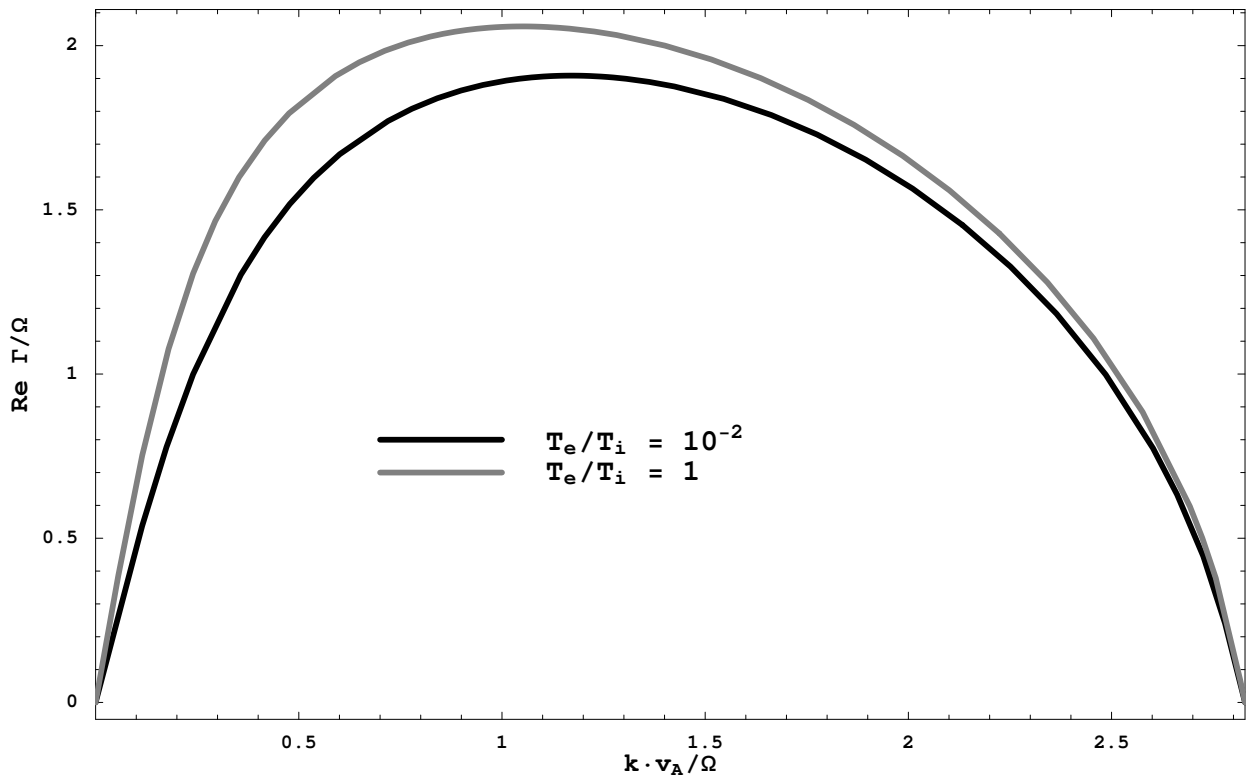}
	\caption{Plot of the real part of the growth rate of the CMVTI, for the case where the ion temperature is 
    much larger than the electron temperature $T_{i0} / T_{e0} = 10^2$, and the case where they are equal. $\alpha_P = 5$, and $\alpha_T = 2$ -- marginal convective stability. This figure, and a
    more comprehensive plasma response incorporating electron pressure dynamics and finite equilibrium electron
    temperature, is taken from \cite{Islam07}.} \label{fig:no_change_include_electron_temperature}
\end{figure}

Here we consider an equilibrium density and temperature distribution given in \S\ref{sec:nomenclature}. Assume axisymmetric perturbations to equilibrium quantities of the form $\delta a \propto \exp\paren{ik_R R + ik_Z z + \Gamma t}$, and define $k_{\|} = {\bf k}\cdot{\bf b}_0$. Eq.~(\ref{eq:drift-kinetic}) then reduces to the following form for ions and electrons, where we assume equal scale heights of radial and vertical ion and electron temperature gradients:
\begin{eqnarray}
  &&\begin{aligned}
      &\delta f_i / f_{i0} = \frac{m_i v_{\|}}{k_B
        T_{i0}}\paren{\frac{-ik_{\|}\mu\delta B + e\delta
          E_{\|}/m_i}{\Gamma + ik_{\|}v_{\|}} - 
        \frac{\paren{2\Omega + \Omega'R}\Gamma  +
          ik_{\|} v_{\|}\Omega'R }{ik_{\|}\paren{\Gamma +
            ik_{\|}v_{\|}}}\bar{B}_R \sin\chi} - \\  
      &\frac{\bar{B}_R}{ik_{\|}}\paren{\parti{\ln n_0}{R} -
        \frac{3}{2}\parti{\ln T_0}{R} + \paren{\frac{m_i\mu B_0}{k_B T_{i0}} + \frac{m_i v_{\|}^2}{2 k_B T_{i0}}}\parti{\ln T_{0}}{R}} + \frac{\bar{B}_R
        v_{\|}\partial\ln p_0/\partial R}{\Gamma + ik_{\|} v_{\|}},
    \end{aligned} \label{eq:perturbed_ion_distribution_function} \\
  &&\begin{aligned}
      &\delta f_e / f_{e0} = \frac{m_e v_{\|}}{k_B
        T_{e0}}\paren{\frac{-ik_{\|}\mu\delta B - e\delta
          E_{\|}/m_e}{\Gamma + ik_{\|}v_{\|}} - 
        \frac{\paren{2\Omega + \Omega'R}\Gamma +
          ik_{\|} v_{\|}\Omega'R}{ik_{\|}\paren{\Gamma +
            ik_{\|}v_{\|}}}\bar{B}_R \sin\chi} - \\ 
      &\frac{\bar{B}_R}{ik_{\|}}\paren{\parti{\ln n_0}{R} -
        \frac{3}{2}\parti{\ln T_0}{R} + \paren{\frac{m_e\mu B_0}{k_B
            T_{e0}} + \frac{m_e v_{\|}^2}{2 k_B T_{e0}}}\parti{\ln
          T_0}{R}} + \frac{\bar{B}_R
        v_{\|}\partial\ln p_0/\partial R}{\Gamma + ik_{\|} v_{\|}}.
    \end{aligned} \label{eq:perturbed_electron_distribution_function}
\end{eqnarray}
Terms with $\Omega$ arise due to the fact that the plasma is rotating; terms with equilibrium gradients of temperature, density, or pressure may drive convective and free energy gradient instabilities. $\delta E_{\|}$ is the electric field that ensures quasineutrality, i.e. $\int \delta f_i^0 B\,d\mu = \int \delta f_e^0 B\,d\mu$. One can demonstrate that in the limit of dominating ion thermal energy $T_{i0} \gg T_{e0}$ that the electric field $\delta E_{\|}$ and electron dynamic terms (such as $\delta p_{e\perp,\|}$) become unimportant in describing the plasma dynamics. This is the simplification employed by \cite{Quataert02c} and \cite{Sharma03}. However, with equilibrium electron temperatures up to one-tenth that of the ion temperatures, as implied by local nonlinear simulations of the collisionless MRI \citep{Sharma07}, the CMVTI dispersion relation is not substantially altered. Fig.~(\ref{fig:no_change_include_electron_temperature}) shows that the dispersion relation of the CMVTI is not significantly different between cases where the electron temperature is negligible ($T_{e0} = 10^{-2} T_{i0}$) and where the electron temperature equals the ion temperature.

Using the induction equation Eq.~(\ref{eq:induct_total}) and the continuity equation, the total force balance equation, Eq.~(\ref{eq:MHD_force_balance}), is represented by the following in terms of Eq.~(\ref{eq:normalization_variables}):
\begin{eqnarray}
  &&\begin{aligned}{}
      &\gamma^2\bar{\bf B} - \gamma^2{\bf
        b}_0\paren{\frac{\delta\rho}{\rho} - \frac{\alpha_P - \alpha_T}{
          i x \beta^{1/2}}\bar{B}_R} + 2\di{\ln\Omega}{\ln
        R}\bar{B}_R\Rvec + 2\gamma \sin\chi\paren{\frac{\delta\rho}{\rho}
        - \frac{\alpha_P - \alpha_T}{i x \beta^{1/2}}\bar{B}_R}\Rvec + \\
      &2\gamma\zvec\times\bar{\bf B} = 
      \hat{\bf k}x\beta\frac{\delta
        p_{\perp}}{p_0} + x^2\beta\frac{\delta p_{\|} - \delta
        p_{\perp}}{p_0}{\bf b}_0 - ix\beta^{1/2}\alpha_P
      \frac{\delta\rho}{\rho}\Rvec - x^2\bar{\bf B} + \hat{\bf k}x
      \frac{\delta B}{B},
    \end{aligned} \label{eq:perturbed_force_balance}
\end{eqnarray}
$\delta B/B = \bar{B}_{\phi} \sin\chi - \paren{k_R/k_Z} \bar{B}_R \cos\chi$, $\delta p_{\|} = \delta p_{i\|} + \delta p_{e\|}$, and $\delta p_{\perp} = \delta p_{i\perp} + \delta p_{e\perp}$. Contributions due to $\delta\rho/\rho - \paren{\alpha_P - \alpha_T}/\paren{i x \beta^{1/2}}\bar{B}_R$ arise from finite plasma compressibility; in the Boussinesq limit these terms are set to zero. The eigenvalue problem consists of three equations for solving $\bar{B}_R$, $\bar{B}_{\phi}$, and $\delta\rho/\rho$: radial force balance, azimuthal force balance, and force balance along ${\bf b}_0$.
\begin{eqnarray}
  &&\begin{aligned}
      &\paren{\gamma^2 + x^2\brak{1 + \frac{k_R^2}{k_Z^2}} +
        2\di{\ln\Omega}{\ln R} - 2\gamma \sin\chi \frac{\alpha_P -
          \alpha_T}{i x \beta^{1/2}}}\bar{B}_R - \paren{2\gamma +
        x^2 \tan\chi \frac{k_R}{k_Z}}\bar{B}_{\phi} + \\
      &\frac{\delta\rho}{\rho}\paren{2\gamma \sin\chi + i x
        \beta^{1/2}\alpha_P} = \frac{k_R}{k_Z \cos\chi} x^2\beta
      \frac{\delta p_{\perp}}{p_0},
    \end{aligned} \label{eq:perturbed_radial_force_balance} \\
  &&\begin{aligned}{}
      &\paren{\gamma^2 \sin\chi\frac{\alpha_P - \alpha_T}{i x
          \beta^{1/2}} + 2\gamma}\bar{B}_R + \paren{\gamma^2 +
        x^2}\bar{B}_{\phi} - \gamma^2 \sin\chi\frac{\delta\rho}{\rho}
      = x^2 \beta \frac{\delta p_{\|} - \delta p_{\perp}}{p_0}\sin\chi,
    \end{aligned} \label{eq:perturbed_azimuthal_force_balance} \\
  &&\begin{aligned}{}
      &\bar{B}_R\paren{\gamma^2 \frac{\alpha_P - \alpha_T}{i x
          \beta^{1/2}} - \gamma^2 \frac{k_R}{k_Z} \cos\chi + 2\gamma
        \sin\chi} + \gamma^2 \sin\chi \bar{B}_{\phi} - \gamma^2
      \frac{\delta\rho}{\rho} = x^2 \beta 
      \frac{\delta p_{\|}}{p_0},
    \end{aligned} \label{eq:perturbed_magnetic_force_balance}
\end{eqnarray}
$\delta p_{\perp}$ and $\delta p_{\|}$ are linear functions of $\bar{B}_R$, $\bar{B}_{\phi}$, and $\delta\rho/\rho$. In subsequent subsections we explore the dispersion relation associated with the rotational magnetothermal and magnetoviscous instabilities. We work in the limit of small electron thermal energies. Therefore, subsequent expressions for perturbed and equilibrium pressure will refer to the ionic component (e.g., $\delta p_{i,\perp} \to \delta p_{\perp}$, $p_{i0} \to p_0$, $p_{i0} \to p_0$, $T_{i0} \to T_i$). 


\subsection{Expressions For Perturbed Pressure} \label{subsec:perturbed_pressure_disp_relation}
\begin{figure}
  \includegraphics[width=\linewidth]{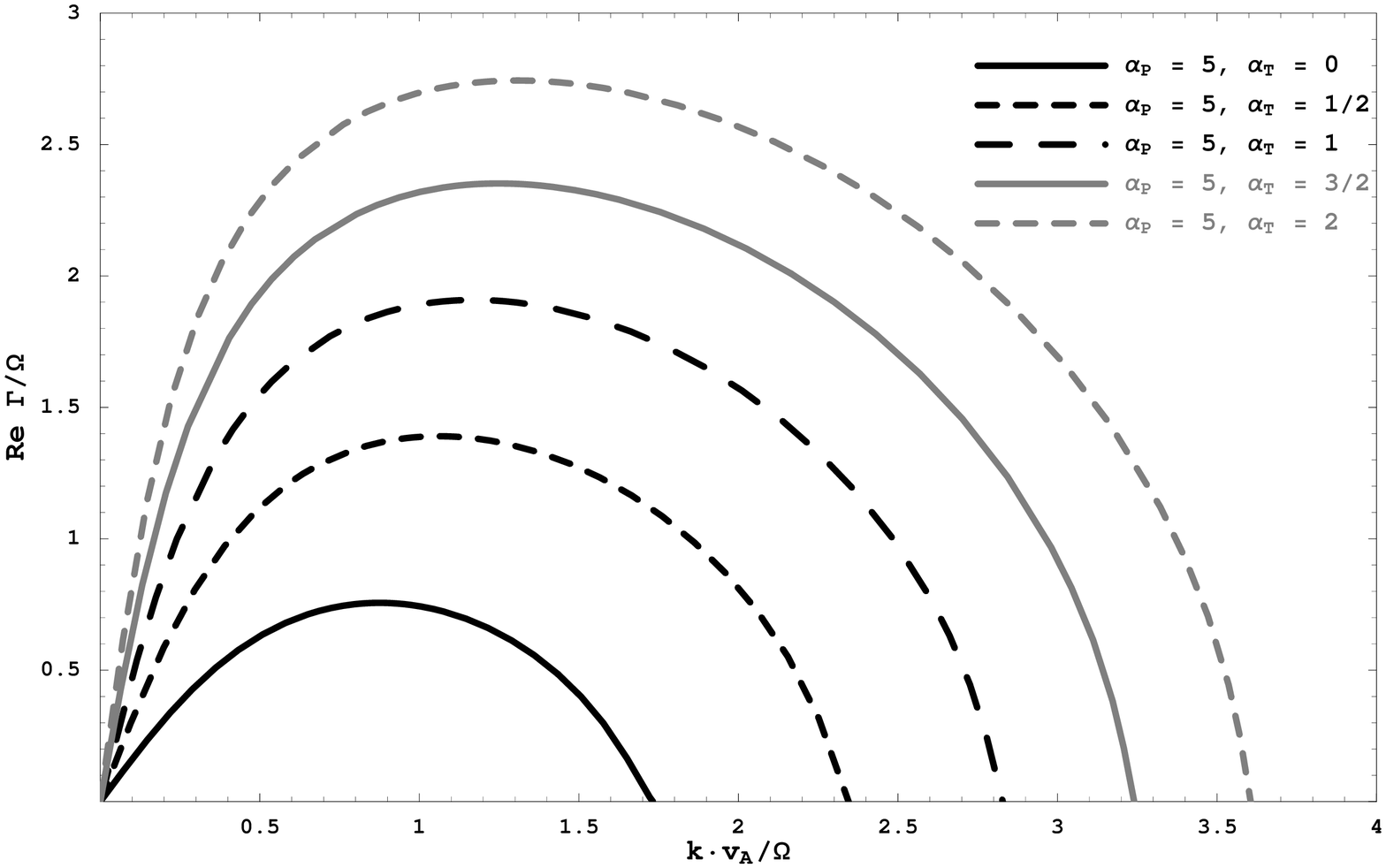}
  \caption{Plot of the real part of the growth rate for the CMVTI and different equilibrium radial temperature gradients. Here $\alpha_P = 5$ and different $\alpha_T = 0$, such that $0 < \alpha_T < \frac{2}{5}\alpha_P$, so that the plasma remains convectively stable.} \label{fig:growthrate_real_fullsimplified}
\end{figure}

In this section we derive expressions for the perturbed parallel and perpendicular pressures, used in closing the eigenmodal equations for the CMVTI (Eqs.~[\ref{eq:perturbed_radial_force_balance}, \ref{eq:perturbed_azimuthal_force_balance}, \ref{eq:perturbed_magnetic_force_balance}]). From Eq.~(\ref{eq:perturbed_ion_distribution_function}), expressions for perturbed parallel and perpendicular pressure can be simplified into a linear combination of density, $\delta B/B$, and $\bar{B}_R$,
\begin{eqnarray}
  &&\frac{\delta p_{\perp}}{p_{0}} = \frac{\delta\rho}{\rho} -
  \frac{\delta B}{B}\paren{R\paren{i\zeta} - 1} +
  \frac{\bar{B}_R}{ik_{\|}}\paren{\parti{\ln n_0}{R} - \parti{\ln
      p_{0}}{R}}, \label{eq:perpendicular_pressure} \\
  &&\begin{aligned}
      \frac{\delta p_{\|}}{p_{0}} =& \paren{\frac{1 - 2\zeta^2
          R\paren{i\zeta}}{R\paren{i\zeta}}}\frac{\delta\rho}{\rho} -
      \paren{\frac{1 - \brak{1 +
            2\zeta^2}R\paren{i\zeta}}{R\paren{i\zeta}}}\frac{\delta
        B}{B} + \\
        &\frac{\bar{B}_R}{ik_{\|}}\paren{\frac{1 - 2\zeta^2
          R\paren{i\zeta}}{R\paren{i\zeta}}\times\parti{\ln n_0}{R}
        - \parti{\ln p_{0}}{R}}.
    \end{aligned} \label{eq:parallel_pressure}
\end{eqnarray}
$\zeta = \Gamma/\paren{k_{\|} \theta_0 \sqrt{2}}$ and $R\paren{\zeta}$ is the plasma response function,
\begin{eqnarray}
  &&R\paren{\zeta} = \frac{1}{\sqrt{\pi}} \int_{-\infty}^{\infty}
  \frac{x e^{-x^2}}{x - \zeta}\,dx. \label{eq:plasma_response}
\end{eqnarray}
Since the phase velocity of the modes are at best of order the sound speed, i.e. $\abs{\zeta} \ltsim 1$, these
perturbations are not adiabatic and the opposite, slow wave ($\abs{\zeta} \ll 1$) limit, holds for most unstable
wavenumbers. The plasma response function in the slow wave limit is,
\begin{eqnarray}
  &&R\paren{i\zeta} = 1 - \zeta\sqrt{\pi} + {\mathcal O}\paren{\zeta^2}. \label{eq:R_low_limit}
\end{eqnarray}
Expressions for perturbed pressure reduce to the following, to first order in $\zeta$:
\begin{eqnarray}
  &&\frac{\delta p_{\|}}{p_{0}} \to \frac{\delta\rho}{\rho} + \sqrt{\pi}\zeta \frac{\delta B}{B} - \xi_R\parti{\ln
  T_0}{R}, \label{eq:slow_limit_parallel_pressure} \\
  &&\frac{\delta p_{\perp}}{p_{0}} \to \frac{\delta\rho}{\rho} - \sqrt{\pi}\zeta \frac{\delta B}{B} +
  \xi_R\paren{3\parti{\ln n_0}{R} - \parti{\ln p_{0}}{R}}. \label{eq:slow_limit_perp_pressure}
\end{eqnarray}
From the radial component of Eq.~(\ref{eq:induct_total}), $\bar{B}_R = ik_{\|}\xi_R$, where $\xi_R$ is the radial fluid displacement. Dispersion relations for the CMVTI are displayed in Fig.~(\ref{fig:growthrate_real_fullsimplified}).
One feature of the plasma response via the CMVTI is that of relatively strong collisionless Barnes damping of magnetohydrodynamic modes along the magnetic fields for long wavelength modes $k_{\|} < \Omega/ \theta_0^{1/2}$, such that at these wavenumbers the phase velocity remains of the order of the sound speed. This feature has been noted in previous studies of the collisionless MRI \citep{Quataert02c, Sharma03}. This damping has the effect of suppressing pressure variations for sufficiently small wavelengths. As the equilibrium plasma $\beta$ decreases to order 1 and smaller the effects of anisotropic pressure become insignificant over much of the range of unstable wavenumbers. Dispersion relations for the CMVTI are similar to the MVTI \citep{Islam12}. The range of unstable wavenumbers match between fluid and collisionless analogues:
\begin{eqnarray}
	&&0 \le k^2 v_A^2 / \Omega \le 2\abs{\di{\ln\Omega}{\ln R}} + \alpha_P \alpha_T. \label{eq:range_unstable_wavenumbers}
\end{eqnarray}
Instead of collisionless damping in the case of the instabilities analyzed within this paper, in fluid treatments it is finite (but dynamically important) viscosity and thermal conductivity that plays this role. Fig.~1 from \cite{Islam12}, showing the real part of the dispersion relation for the MVTI for a dynamically important viscous diffusion coefficient, realistic Prandtl number, for a range of convectively stable equilibria.
\begin{figure}[!ht]
	\includegraphics[width=\linewidth]{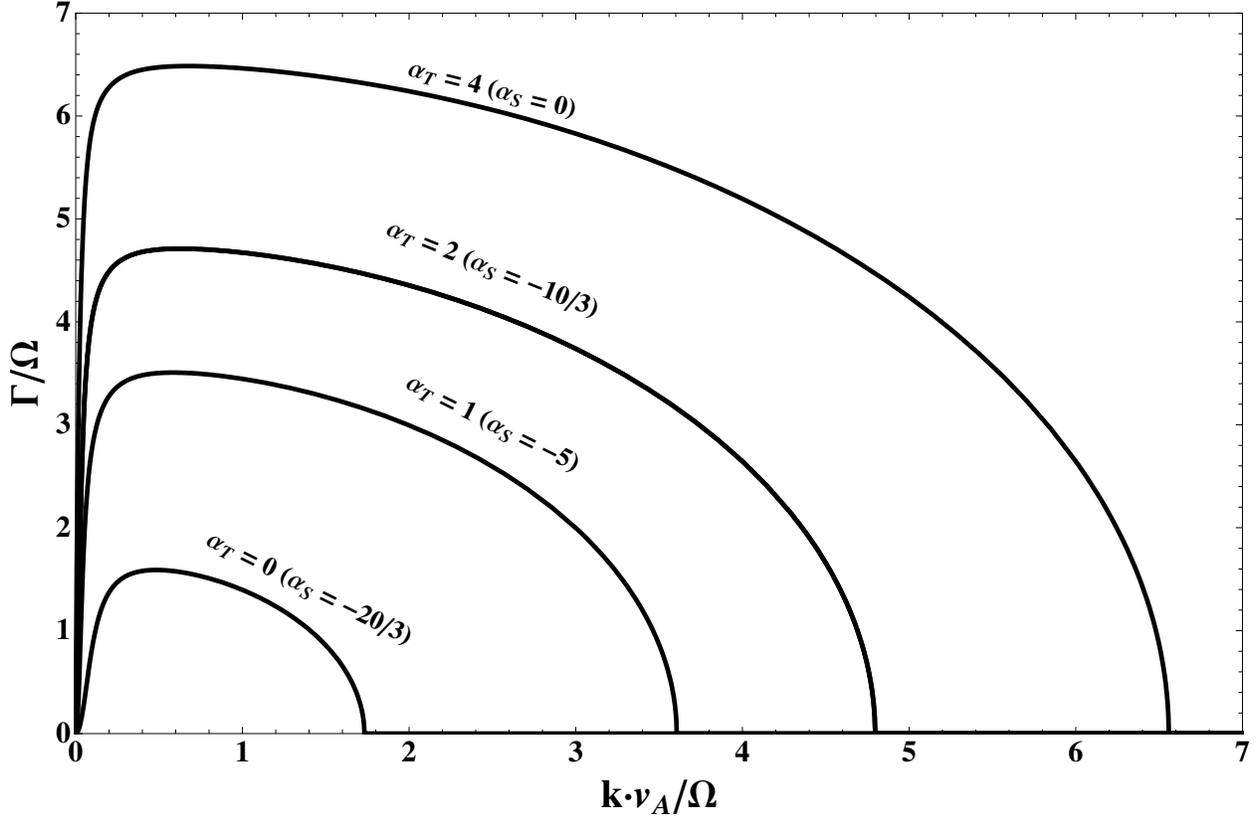}
	\caption{Plot of the real portion of the growth rate for th MVTI various $\alpha_T$. $\alpha_P = 10$, viscous diffusion coefficient $\nu\Omega/v_A^2 = 10^2$, Prandtl number $\text{Pr} = 1/101$, and $\alpha_S = 5\alpha_T / 3 - 2\alpha_P / 3$. Rollover occurs at wavenumbers $k \sim \sqrt{\Omega/\nu} \ll \Omega / v_A$.} \label{fig:mvi}
\end{figure}
Efficient viscosity and thermal diffusivity dissipates the MVTI at wavelengths such that $\nu k^2 \ltsim \Omega$, where $\nu$ is a viscous diffusion coefficient along magnetic field lines \citep{Islam2005, Islam12}. Fluid simulations that model collisionless damping as a form of fluid transport \citep{Sharma06, Sharma07}, do so with heat fluxes whose thermal diffusion coefficients are on the order of $\theta_0 / \Omega$.

\subsection{Quadratic Fluxes} \label{subsec:quadraticfluxes}
\begin{figure}[!ht]
	\includegraphics[width =\linewidth]{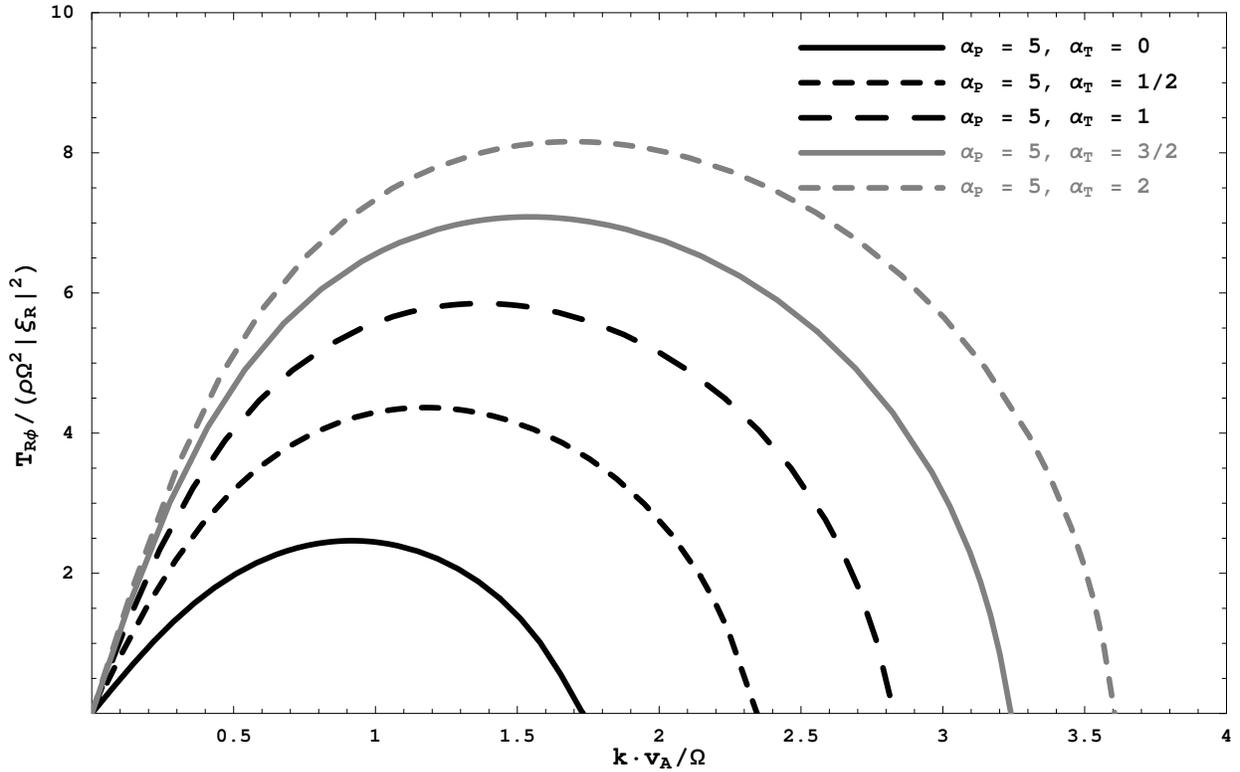}
	\caption{Outwards normalized azimuthal stress for the CMVTI and various convectively
	stable equilibrium profiles with $\alpha_P = 5$ and $0 \le \alpha_T \le 2$.} \label{fig:MVTI_collisionless_reynoldsstress}
\end{figure}

\begin{figure}[!ht]
  \includegraphics[width = \linewidth]{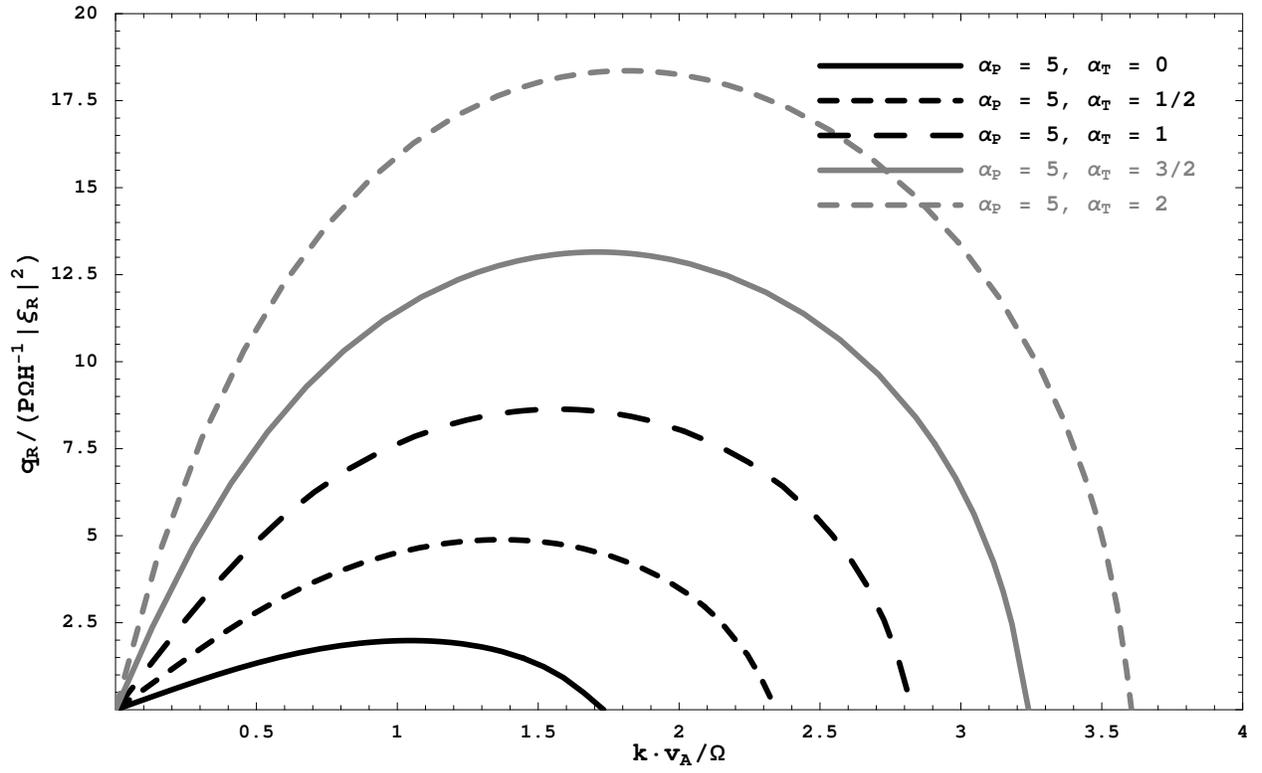}
  \caption{Same as Fig.~(\ref{fig:MVTI_collisionless_reynoldsstress}), except for
    quadratic heat flux.} \label{fig:MVTI_collisionless_heatflux}
\end{figure}

\begin{figure}[!ht]
  \includegraphics[width =
  \linewidth]{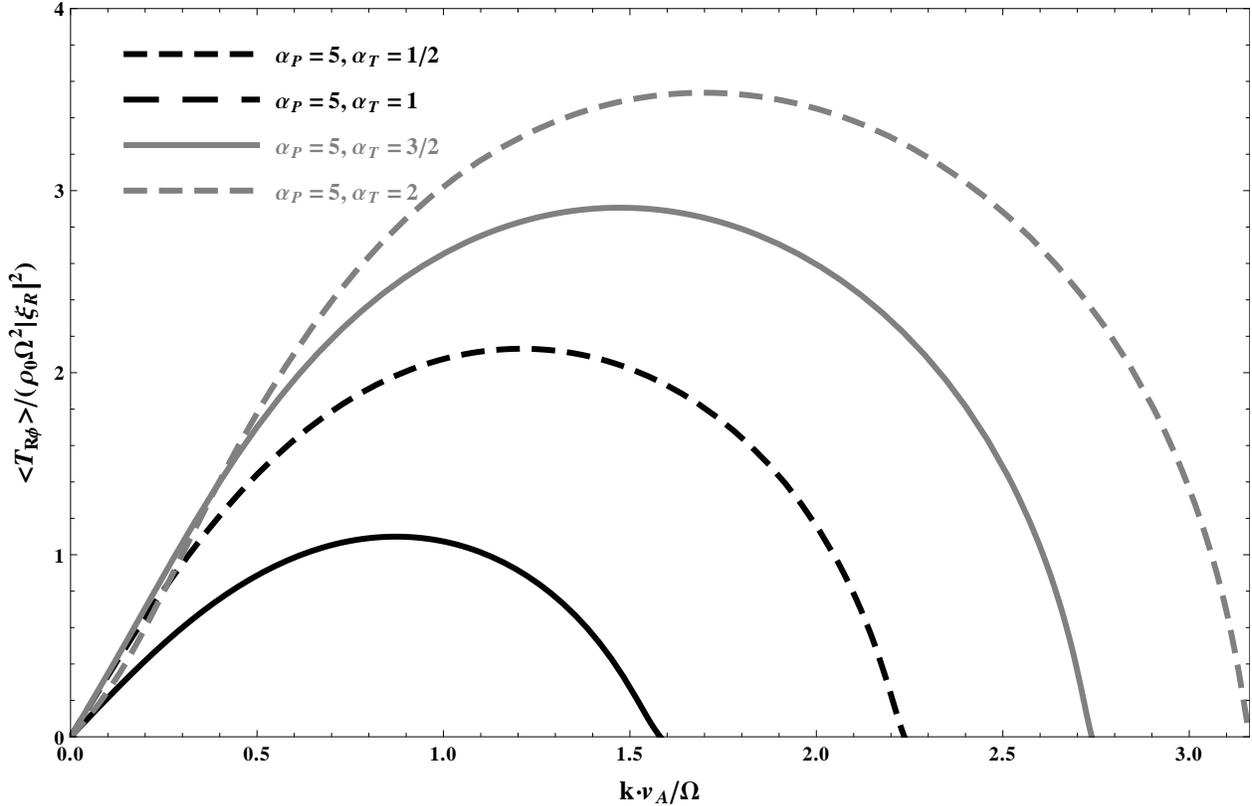}
  \caption{Plot of the azimuthal stress for the CMVTI and zero rotational shear, and various convectively stable equilibrium profile. } \label{fig:MVTI_collisionless_reynoldsstress_zeroshear}
\end{figure}

\begin{figure}[!ht]
  \includegraphics[width=\linewidth]{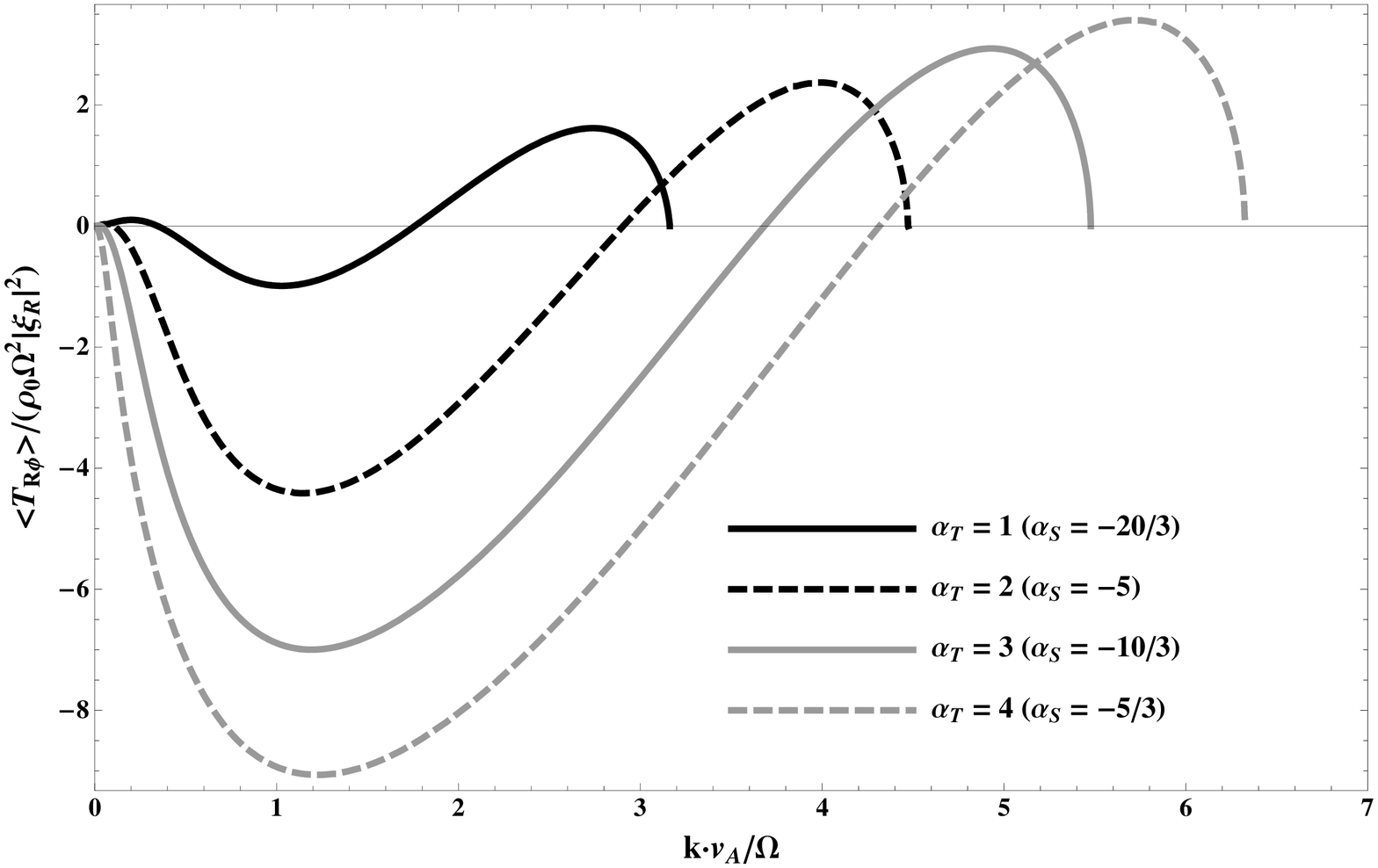}
  \caption{Normalized flux $\braket{T_{R\phi}}/\paren{\rho\Omega^2\abs{\xi_R}^2}$ for a rigid rotation profile ($\Omega'R = 0$), and convectively stable equilibria, for the MVTI. $\alpha_S = 5\alpha_T / 3 - 2\alpha_P / 3$.} \label{fig:MTI_angmomflux_rigidrotation}
\end{figure}

Here, we determine the normalized quadratic heat flux, Eq.~(\ref{eq:averaged_radial_heat_flux}), and the radial azimuthal stress, Eq.~(\ref{eq:azimuthal_stress_average}), associated with a given mode of purely vertical wavenumber $k_Z$. We normalize these fluxes as a function of fixed Lagrangian radial displacement $\xi_R = \delta u_R/\Gamma$. From Eq.~(\ref{eq:perturbed_ion_distribution_function}), we have the following expressions for $\delta u_{\|}$, $\delta q_{\|}$, and $\delta q_{\perp}$.
\begin{eqnarray}
  &&\begin{aligned}
      &\frac{\delta u_{\|}}{\theta_0^{1/2}} = -i\zeta\sqrt{2} R\paren{i\zeta}
      \paren{\frac{\delta B}{B} -
        \frac{2\Omega\Gamma}{k_{\|}^2 \theta_0}\bar{B}_R \sin\chi +
        \frac{i\bar{B}_R}{k_{\|}}\paren{\parti{\ln p_{i0}}{R}}} + 
      \frac{i\bar{B}_R}{k_{\|}}\sin\chi\Omega'R,
    \end{aligned} \label{eq:perturbed_parallel_velocity} \\
  &&\begin{aligned}
      &\frac{\delta q_{\|}}{p_{0} \theta_0^{1/2}} = \paren{i\zeta\sqrt{2}\paren{
          \frac{\delta B}{B}} - \frac{\bar{B}_R}{k_{\|}}
        \paren{\parti{\ln p_{0}}{R}}\zeta\sqrt{2} - 
        \frac{\Omega\zeta^2}{k_{\|}
          \theta_0^{1/2}}i \bar{B}_R\cos\chi}\paren{\brak{2\zeta^2 + 3}
        R\paren{i\zeta} - 1},
    \end{aligned} \label{eq:heatflux_parallel_perturbed} \\
  &&\begin{aligned}
      &\frac{\delta q_{\perp}}{p_{0} \theta_0^{1/2}} = -i\zeta\sqrt{2}\paren{\frac{\delta
          B}{B}}R\paren{i\zeta}.
    \end{aligned} \label{eq:heatflux_perp_perturbed}
\end{eqnarray}
Expressions for the heat flux and radial-azimuthal stress for these axisymmetric modes at the disk midplane are given by the following:
\begin{eqnarray}
  &&W_{R\phi} = \text{Re}\paren{\rho_0 \delta u_R^* \delta u_{\phi} -
    v_A^2\bar{B}_R^* \bar{B}_{\phi} + \sin\chi\bar{B}_R^*
    \delta p_v}, \label{eq:azimuthal_stress_modal} \\
  &&F_{ER} = \text{Re}\paren{\frac{5}{2}\delta u_R^* \delta\theta -
    \delta q \bar{B}_R^* - \frac{1}{3}\delta p_v
    \bar{B}_R^*}. \label{eq:heat_flux_modal}
\end{eqnarray}
One can employ expressions for the total pressure (Eq.~[\ref{eq:total_pressure}]), pressure difference (Eq.~[\ref{eq:pressure_difference}]), and total heat flux (Eq.~[\ref{eq:heat_flux}]) with
expressions for the perturbed pressures (Eqs.~[\ref{eq:perpendicular_pressure}] and [\ref{eq:parallel_pressure}]) and heat fluxes given (Eqs.~[\ref{eq:heatflux_parallel_perturbed}] and [\ref{eq:heatflux_perp_perturbed}]). The form of the relative perturbed density and toroidal magnetic field $\delta\rho/\rho$ and $\bar{B}_{\phi}$ are described in the eigenvalue equations (Eqs.~[\ref{eq:perturbed_radial_force_balance}], [\ref{eq:perturbed_azimuthal_force_balance}], and [\ref{eq:perturbed_magnetic_force_balance}]). Using variable scalings as given by Eq.~(\ref{eq:normalization_variables}), expressions for $\delta u_R$, $\bar{B}_R$, $\bar{B}_{\phi}$, $\delta u_{\phi}$, $\delta p_v$, $\delta\theta$, and $\delta q$ in terms of $\xi_R$ are,
\begin{eqnarray}
  &&\delta u_R = \gamma\paren{\Omega\xi_R}, \label{eq:delta_uR} \\
  &&\bar{B}_R = ix\paren{\frac{\Omega}{v_A}\xi_R}, \label{eq:bar_BR} \\
  &&\begin{aligned}
      \bar{B}_{\phi} =& -\frac{%
        2\gamma\paren{\cos^2\chi +
          R\paren{\frac{i\gamma}{x\sqrt{2\beta}} \sin^2\chi}} - 
        i x \beta^{1/2} \alpha_P \sin\chi
        \brak{R\paren{\frac{i\gamma}{x\sqrt{2\beta}}} - 1}}{%
        \gamma^2\paren{\cos^2\chi +
          R\paren{\frac{i\gamma}{x\sqrt{2\beta}}}\sin^2\chi} + x^2 - 
        2x^2\beta \sin^2\chi\brak{\paren{1 + \frac{\gamma^2}{2x^2\beta}}
          R\paren{\frac{i\gamma}{x\sqrt{2\beta}}} - 1}}\times\\
      &ix\paren{\frac{\Omega}{v_A}\xi_R}, \end{aligned} \label{eq:bar_Bphi}
  \\
  &&\begin{aligned}
      \delta u_{\phi} =& \frac{\gamma}{ix}v_A\bar{B}_{\phi}
      \paren{\cos^2\chi + R\paren{\frac{i\gamma}{x\sqrt{2\beta}}}b_{\phi
          0}^2} + \\
      &\paren{\Omega\xi_R}\paren{\abs{\di{\ln\Omega}{\ln R}} -
        \sin\chi\brak{\frac{2\gamma^2}{x^2\beta}\sin\chi +
          i\alpha_P \frac{\gamma}{x\beta^{1/2}}}
        R\paren{\frac{i\gamma}{x\sqrt{2\beta}}}},
    \end{aligned} \label{eq:delta_uphi} \\
  &&\begin{aligned}{}
      \delta p_v =& \paren{p_{0} H^{-1} \xi_R}
      \paren{\brak{\frac{\gamma^2}{x^2\beta} - 1}R\paren{
          \frac{i\gamma}{x\sqrt{2\beta}}} + 1} + \\
      &2\paren{p_{0}\bar{B}_{\phi} \sin\chi - i \beta^{-1}
        \frac{\gamma}{x} p_{0} \xi_R \frac{\Omega}{v_A}}\paren{
        \brak{1 + \frac{\gamma^2}{2x^2\beta}}
        R\paren{\frac{i\gamma}{x\sqrt{2\beta}}} - 1},
    \end{aligned} \label{eq:delta_pressure_viscous} \\
  &&\begin{aligned}
      \frac{\delta\theta}{\theta_0} =& \frac{\delta p}{p_{0}} -
      \frac{\delta\rho}{\rho} = \paren{\xi_R H^{-1}}\paren{\alpha_T +
        \alpha_P\paren{\brak{\frac{5}{3} +
            \frac{\gamma^2}{3x^2\beta}}
          R\paren{\frac{i\gamma}{x\sqrt{2\beta}}} - \frac{5}{3}}} + \\
      &\frac{1}{3}\bar{B}_{\phi} \sin\chi\paren{
        \brak{\frac{\gamma^2}{2x^2\beta} - 1} R\paren{\frac{i\gamma}{
            x\sqrt{2\beta}}} + 1} - \\
      &\frac{2}{3}i\beta^{-1} \frac{\gamma}{x} \sin\chi\paren{%
        \frac{\Omega}{v_A}\xi_R}\paren{\brak{1 + 
          \frac{\gamma^2}{x^2\beta}}
        R\paren{\frac{i\gamma}{x\sqrt{2\beta}}} - 1},
    \end{aligned} \label{eq:delta_theta} \\
  &&\begin{aligned}{}
      \delta q =& \paren{p_{0} \Omega\xi_R} \sin\chi
      \paren{i\alpha_P \frac{\gamma}{x\beta^{1/2}} +
        \frac{2\gamma^2}{x^2\beta}}\paren{\brak{\frac{3}{2} +
          \frac{\gamma^2}{2x^2\beta}}
        R\paren{\frac{i\gamma}{x\sqrt{2\beta}}} - \frac{1}{2}} + \\
      &i\paren{p_{0} \theta_0^{1/2}
        \bar{B}_{\phi}}\frac{\gamma}{2x\beta^{1/2}} \sin\chi\paren{%
        \brak{1 + \frac{\gamma^2}{x^2\beta}}
        R\paren{\frac{i\gamma}{x\sqrt{2\beta}}} - 1}.
    \end{aligned} \label{eq:delta_heatflux}
\end{eqnarray}
The azimuthal stress is normalized in units of $\rho_0 \Omega^2\abs{\xi_R}^2$ and the heat flux in terms of $\rho_0 \theta_0^{1/2} \Omega^2 \abs{\xi_R}^2 \equiv p_0 \Omega H^{-1} \abs{\xi_R}^2$. The relatively involved quadratic expressions for angular momentum and heat flux are not shown. In Figs.~(\ref{fig:MVTI_collisionless_reynoldsstress}) and (\ref{fig:MVTI_collisionless_heatflux}) are plots of the heat flux and azimuthal stress for the CMVTI for different $0 < \alpha_T < \frac{2}{5}\alpha_P$.
\FloatBarrier
One can easily demonstrate, by setting $\alpha_P = \alpha_T = 0$, that the heat flux for the collisionless MRI is zero. There are no equilibrium radial gradients of temperature or density, the growth rate is purely real, so that for a given mode the temperature and viscous pressure perturbations are out of phase with the perturbed radial velocity, and the perturbed heat flux is out of phase with the perturbed radial magnetic field. The salient features of these instabilities is that they produce the right type of azimuthal stress that can drive accretion.  The general sense of the Reynolds stress is outwards for all unstable wavenumbers for the CMVTI; however, \cite{Islam12} demonstrates that the MVTI can have a generally small range of small wavenumbers for even an unstable Keplerian rotational profile in which the azimuthal stress is negative. Finally, even in the absence of rotational shear $\Omega'R = 0$ the effects of a heat flux can also transport angular momentum outwards; this is demonstrated in Fig.~(\ref{fig:MVTI_collisionless_reynoldsstress_zeroshear}).
Surprisingly, the CMVTI, even in the absence of differential rotation, is more effective at transporting angular momentum outwards than the MVTI. For comparison, Fig.~(\ref{fig:MTI_angmomflux_rigidrotation}) reproduces Fig.~7 from \cite{Islam12}, and demonstrates that for a substantial portion of unstable wavenumbers, the modal MVTI Reynolds stress is inwards rather than outwards.
As noted in \cite{Islam12}, in the absence of rotational shear, no energy can be extracted from the flow (see Eq.~(\ref{eq:averaged_total_energy_equation})). Second, the ambiguity of angular transport for the CMVTI is analogous to the MVTI, in that the CMVTI acts as a mechanism to transport thermal energy outwards, largely (and, in the case of rigid rotation, completely) independent of the manner in which it transports angular momentum. A first step to understand the CMVTI would be to explore unambiguous measures of turbulent, saturated angular momentum and heat flux in local simulations that are marginally stable to the CMVTI and MVTI.

\section{Summary of Results and Further Work} \label{sec:summary_and_future_work}
In this paper we have derived the drift kinetic equation explicitly in a rotating frame with possible significant gas pressures and only mild collisionality, with application to hot, dilute, weakly-magnetized (in the sense that magnetic forces are subdominant in equilibrium), at best mildly relativistic systems such as as dim accretion about supermassive black holes. \S\ref{subsec:summary_of_results} describes the main results of this paper. \S\ref{subsec:future_work} elaborates on the main directions for future work.

\subsection{Summary of Results} \label{subsec:summary_of_results}
We see physical terms explicitly associated with disk stratification as well as rotation. We also see that one may rather easily derive modifications of the azimuthal stress and heat flux due to fluctuations or waves in accreting systems \citep{Balbus98, Balbus03} due to dilute plasmas, as demonstrated in \S\ref{sec:dilute_fluxes}, in order to characterize how or whether instabilities may create the right type of turbulence that drives accretion.

We have analysed the CMVTI, which have been demonstrated \citep{Islam2005, Islam12} from a fluid treatment to destabilize a plasma, through anisotropic viscosities and thermal conductivities, that possesses adverse angular velocity or temperature gradients. We demonstrate the congruence in the dispersion relation for the CMVTI with the MVTI. Heat fluxes and azimuthal stresses associated with this instability have the right sense (i.e., positive), to drive accretion in fat dilute nonradiative rotating plasmas, and roughly match their respective fluid counterparts. Furthermore, we note that expressions for the normalized pressure and heat gradients, $\alpha_P$ and $\alpha_T$, go as $H / R$ if we assume that equilibrium temperature and pressure radial scale heights are of order the disk radius. Therefore, we expect only geometrically thick disks to efficiently transport angular momentum in nonradiative accretion flows.

\subsection{Future Work} \label{subsec:future_work}
Although we have applied the drift-kinetic equation to a single but important class of instability in Keplerian-like rotating systems, its representation as given in Eq.~(\ref{eq:drift-kinetic}) lends itself to much richer studies of these types of dilute plasmas. Immediate analytic work can enhance our understanding of the stability of a collisionless nonradiative accretion disk to the CMVTI. Due to the requirement of geometrically thick disks to efficiently transport angular momentum without radiative losses, a global stability analysis with realistic disk structure is needed.

Fluid MHD models of local nonlinear evolution in collisionless astrophysical plasmas have employed prescriptions to model collisionless and fast, small-length scale isotropizing phenomena. First, Landau fluid expressions of heat flux and viscosity  represent, as practical as is possible, the collisionless momentum and heat transport driven by the instabilities of interest. And second, a hard wall on relative ionic pressure anisotropies reflects observations of marginal pressure anisotropy in the solar wind \citep{Hellinger06, Bale09}, due to unresolvable fast (on the order of the ion gyroperiod) and short wavelength (on the order of the ion gyroradius) instabilities driven by pressure anisotropy. Similar pressure anisotropies are found to develop for the CMVTI, as shown by a more comprehensive stability analysis \cite{Islam07}. Although these prescriptions have been fruitfully applied to local simulations of the collisionless MRI \citep{Sharma06, Sharma07} and the buoyancy instability \citep{Kunz12}, a more self-consistent numerical model is desired.

A more productive approach would be to use gyrokinetic or drift-kinetic MHD codes, such as Fokker-Planck \citep{Grandgirard06}, ionic particle in cell \citep{Kolesnikov10, Chen09}, or hybrid PIC \citep{Brecht88} modified such that ions move drift-kinetically, to simulate the dynamics of these plasmas. Recent work in modifying full particle in cell \citep{Riquelme12} and 3D hybrid particle in cell \citep{Kunz14} for co-rotating local reference frames has found promise in the study of the collisionless differentially rotating plasmas, currently under situations in which the separation of length and time scales with ion gyromotion, disk rotational frequency, and the fastest growing wavelengths of the MRI are not too severe. These numerical models have shown promise in understanding the nonlinear development of initially weak-field (ion gyroradius larger than the wavelength of the fastest growing mode) magnetotational instabilities \citep{Krolik06, Ferraro07}. Enhancements to these codes towards larger spatial and temporal separations between ion gyromotion and the slower, longer scale dynamics of collisionless MHD make them well suited towards understanding the nature of heat flux and angular momentum transport in the CMVTI.

\section{Acknowledgements}
The author would like to thank the referees, whose input has clarified and focused this paper into a generalization of the MVTI into the collisionless regime, for pointing out a crucial reference \citep{Sharma07} demonstrating that the collisionless MRI may heat electrons such that even collisionless accretions flows may become radiative, and for allowing a further iteration to repair mistakes in content.


\begin{thebibliography}{44}
\expandafter\ifx\csname natexlab\endcsname\relax\def\natexlab#1{#1}\fi

\bibitem[{Aitken {et~al.}(2000)Aitken, Greaves, Chrysostomou, Jenness, Holland,
  Hough, Pierce-Price, \& Richer}]{Aitken00}
Aitken, D.~K., Greaves, J., Chrysostomou, A., Jenness, T., Holland, W.~S.,
  Hough, J.~H., Pierce-Price, D., \& Richer, J. 2000, The Astrophysical Journal
  Letters, 534, L173

\bibitem[{Baganoff {et~al.}(2003)Baganoff, Maeda, Morris, Bautz, Brandt, Cui,
  Doty, Feigelson, Garmire, \& Pravdo}]{Baganoff03}
Baganoff, F.~K., Maeda, Y., Morris, M.~R., Bautz, M.~W., Brandt, W.~N., Cui,
  W., Doty, J.~P., Feigelson, E.~D., Garmire, G.~P., \& Pravdo, S.~H. 2003, The
  Astrophysical Journal, 591, 891

\bibitem[{Balbus(2003)}]{Balbus03}
Balbus, S.~A. 2003, Annual Review of Astronomy and Astrophysics, 41, 555

\bibitem[{Balbus(2004)}]{Balbus04a}
---. 2004, The Astrophysical Journal, 600, 865

\bibitem[{Balbus \& Hawley(1991)}]{Balbus91}
Balbus, S.~A. \& Hawley, J.~F. 1991, The Astrophysical Journal, 376, 214

\bibitem[{Balbus \& Hawley(1998)}]{Balbus98}
---. 1998, Reviews of Modern Physics, 70, 1

\bibitem[{Bale {et~al.}(2009)Bale, Kasper, Howes, Quataert, Salem, \&
  Sundkvist}]{Bale09}
Bale, S., Kasper, J., Howes, G., Quataert, E., Salem, C., \& Sundkvist, D.
  2009, Physical Review Letters, 103, 211101

\bibitem[{Bisnovatyi-Kogan \& Seidov(1985)}]{Bisnovatyi-Kogan85}
Bisnovatyi-Kogan, G.~S. \& Seidov, Z.~F. 1985, Astrophysics and Space Science,
  115, 275

\bibitem[{Bower {et~al.}(2003)Bower, Wright, Falcke, \& Backer}]{Bower03}
Bower, G., Wright, M. C.~H., Falcke, H., \& Backer, D. 2003, The Astrophysical
  Journal, 588, 331

\bibitem[{Brecht \& Thomas(1988)}]{Brecht88}
Brecht, S.~H. \& Thomas, V.~A. 1988, Computer Physics Communications, 48, 135

\bibitem[{Chandrasekhar(1960)}]{Chandrasekhar60}
Chandrasekhar, S. 1960, Proceedings of the National Academy of Sciences, 46,
  253

\bibitem[{Chang \& Callen(1992{\natexlab{a}})}]{Chang92a}
Chang, Z. \& Callen, J.~D. 1992{\natexlab{a}}, Physics of Fluids B, 4, 1167

\bibitem[{Chang \& Callen(1992{\natexlab{b}})}]{Chang92b}
---. 1992{\natexlab{b}}, Physics of Fluids B, 4, 1182

\bibitem[{Chen \& Parker(2009)}]{Chen09}
Chen, Y. \& Parker, S.~E. 2009, Physics of Plasmas, 16, 052305

\bibitem[{de~Villiers \& Hawley(2003)}]{DeVilliers03a}
de~Villiers, J.-P. \& Hawley, J.~F. 2003, The Astrophysical Journal, 592, 1060

\bibitem[{de~Villiers {et~al.}(2003)de~Villiers, Hawley, \&
  Krolik}]{DeVilliers03b}
de~Villiers, J.-P., Hawley, J.~F., \& Krolik, J. 2003, The Astrophysical
  Journal, 599, 1238

\bibitem[{Ferraro(2007)}]{Ferraro07}
Ferraro, N.~M. 2007, The Astrophysical Journal, 662, 512

\bibitem[{Fromang {et~al.}(2004)Fromang, de~Villiers, \& Balbus}]{Fromang04a}
Fromang, S., de~Villiers, J.-P., \& Balbus, S.~A. 2004, Astrophysics and Space
  Science, 292, 439

\bibitem[{Grandgirard {et~al.}(2006)Grandgirard, Brunetti, Bertrand, Besse,
  Garbet, Ghendrih, Manfredi, Sarazin, Sauter, Sonnendr{\"u}cker, Vaclavik, \&
  Villard}]{Grandgirard06}
Grandgirard, V., Brunetti, M., Bertrand, P., Besse, N., Garbet, X., Ghendrih,
  P., Manfredi, G., Sarazin, Y., Sauter, O., Sonnendr{\"u}cker, E., Vaclavik,
  J., \& Villard, L. 2006, Journal of Computational Physics, 217, 395

\bibitem[{Hawley {et~al.}(1996)Hawley, Gammie, \& Balbus}]{Hawley96}
Hawley, J.~F., Gammie, C.~F., \& Balbus, S.~A. 1996, The Astrophysical Journal,
  464, 690

\bibitem[{Hellinger {et~al.}(2006)Hellinger, Tr{\'a}vn{\'\i}{\v c}ek, Kasper,
  \& Lazarus}]{Hellinger06}
Hellinger, P., Tr{\'a}vn{\'\i}{\v c}ek, P., Kasper, J.~C., \& Lazarus, A.~J.
  2006, Geophysical Research Letters, 33, 09101

\bibitem[{Hinton \& Hazeltine(1976)}]{Hinton76}
Hinton, F.~L. \& Hazeltine, R.~D. 1976, Reviews of Modern Physics, 48, 239

\bibitem[{Islam(2007)}]{Islam07}
Islam, T. 2007, PhD thesis, University of Virginia

\bibitem[{Islam(2012)}]{Islam12}
---. 2012, The Astrophysical Journal, 746, 8

\bibitem[{Islam \& Balbus(2005)}]{Islam2005}
Islam, T. \& Balbus, S.~A. 2005, The Astrophysical Journal, 633, 328

\bibitem[{Kolesnikov {et~al.}(2010)Kolesnikov, Wang, Hinton, Rewoldt, \&
  Tang}]{Kolesnikov10}
Kolesnikov, R.~A., Wang, W.~X., Hinton, F.~L., Rewoldt, G., \& Tang, W.~M.
  2010, Physics of Plasmas, 17, 2506

\bibitem[{Krolik \& Zweibel(2006)}]{Krolik06}
Krolik, J.~H. \& Zweibel, E.~G. 2006, The Astrophysical Journal, 644, 651

\bibitem[{Kulsrud(1983)}]{Kulsrud83}
Kulsrud, R.~M. 1983, in Basic Plasma Physics: Selected Chapters, Handbook of
  Plasma Physics, Volume 1, ed. A.~A. Galeev \& R.~N. Sudan, 1--+

\bibitem[{Kulsrud(2005)}]{Kulsrud05}
Kulsrud, R.~M. 2005, {Plasma Physics for Astrophysics}, Princeton Series in
  Astrophysics (Princeton, NJ: Princeton University Press)

\bibitem[{Kunz {et~al.}(2012)Kunz, Bogdanovi{\'c}, Reynolds, \& Stone}]{Kunz12}
Kunz, M.~W., Bogdanovi{\'c}, T., Reynolds, C.~S., \& Stone, J.~M. 2012, The
  Astrophysical Journal, 754, 122

\bibitem[{Kunz {et~al.}(2014)Kunz, Stone, \& Bai}]{Kunz14}
Kunz, M.~W., Stone, J.~M., \& Bai, X.~N. 2014, Journal of Computational Physics

\bibitem[{Marrone {et~al.}(2005)Marrone, Moran, Zhao, \& Rao}]{Marrone05}
Marrone, D.~P., Moran, J.~M., Zhao, J.~H., \& Rao, R. 2005, The Astrophysical
  Journal, 640, 308

\bibitem[{Narayan(2002)}]{Narayan02b}
Narayan, R. 2002, in Lighthouses of the Universe: The Most Luminous Celestial
  Objects and Their Use for Cosmology: Proceedings of the MPA/ESO/MPE/USM Joint
  Astronomy Conference Held in Garching, Harvard-Smithsonian Center for
  Astrophysics, 60 Garden Street, Cambridge, MA 02138, USA, 405

\bibitem[{Narayan {et~al.}(1998)Narayan, Mahadevan, Grindlay, Popham, \&
  Gammie}]{Narayan98a}
Narayan, R., Mahadevan, R., Grindlay, J.~E., Popham, R.~G., \& Gammie, C.~F.
  1998, The Astrophysical Journal, 492, 554

\bibitem[{Ogilvie(1997)}]{Ogilvie97}
Ogilvie, G.~I. 1997, Monthly Notices of the Royal Astronomical Society, 288, 63

\bibitem[{Quataert {et~al.}(2002)Quataert, Dorland, \& Hammett}]{Quataert02c}
Quataert, E., Dorland, W.~D., \& Hammett, G.~W. 2002, The Astrophysical
  Journal, 577, 524

\bibitem[{Riquelme {et~al.}(2012)Riquelme, Quataert, Sharma, \&
  Spitkovsky}]{Riquelme12}
Riquelme, M.~A., Quataert, E., Sharma, P., \& Spitkovsky, A. 2012, The
  Astrophysical Journal, 755, 50

\bibitem[{Sano \& Stone(2002)}]{Sano02}
Sano, T. \& Stone, J.~M. 2002, The Astrophysical Journal, 570, 314

\bibitem[{Sharma \& Hammett(2006)}]{Sharma06}
Sharma, P. \& Hammett, G.~W. 2006, PhD thesis, Princeton Univ., Washington, DC

\bibitem[{Sharma {et~al.}(2003)Sharma, Hammett, \& Quataert}]{Sharma03}
Sharma, P., Hammett, G.~W., \& Quataert, E. 2003, The Astrophysical Journal,
  596, 1121

\bibitem[{Sharma {et~al.}(2007)Sharma, Quataert, \& Stone}]{Sharma07}
Sharma, P., Quataert, E., \& Stone, J.~M. 2007, The Astrophysical Journal, 671,
  1696

\bibitem[{Velikhov(1959)}]{Velikhov59}
Velikhov, E.~P. 1959, Zhur Eksptl' i Teoret Fiz, 36, 1398

\bibitem[{Wardle(1999)}]{Wardle99}
Wardle, M. 1999, Monthly Notices of the Royal Astronomical Society, 307, 849

\end{thebibliography}
\end{document}